\documentclass[a4paper]{jpconf}
\usepackage{graphicx}
\usepackage{subfig}
\usepackage{wrapfig}
\begin{document}
\title{Track Reconstruction and b-Jet Identification for the ATLAS Trigger System}

\author{Andrea Coccaro, on behalf of the ATLAS Collaboration}

\address{Department of Physics, University of Genoa and I.N.F.N., Genoa, Italy}

\ead{andrea.coccaro@ge.infn.it}

\begin{abstract}
A sophisticated trigger system, capable of real-time track reconstruction, is used in the ATLAS experiment to select interesting events in the proton-proton collisions at the Large Hadron Collider at CERN. A set of $b$-jet triggers was activated in ATLAS for the entire 2011 data-taking campaign and successfully selected events enriched in jets arising from heavy-flavour quarks. Such triggers were demonstrated to be crucial for the selection of events with no lepton signature and a large jet multiplicity. An overview of the track reconstruction and online $b$-jet selection with performance estimates from data is presented in these proceedings.
\end{abstract}

\section{Introduction}
The Large Hadron Collider (LHC) has just finished a tremendously successful year of proton-proton collisions with a total integrated luminosity in 2011 of nearly 6\,fb$^{-1}$ and a peak instantaneous luminosity constantly increasing over the year and reaching its maximum at $3.65\times10^{33}$\,cm$^{-2}$\,s$^{-1}$. The enormous collision rate required an efficient and flexible trigger system to select only a few hundred of events per second and to evolve following the continuous improvement of the LHC performance during the year.

The ability to distinguish between jets arising from heavy-flavour quarks ($b$- and $c$-jets) and light jets (jets from $u$-, $d$-, $s$- and gluon jets) already at the trigger level offers significant improvement and increases the flexibility in the triggering strategy to help maximize the output for physics analyses. The strategy adopted in ATLAS is reviewed in these proceedings. A description of the $b$-jet trigger selection together with details on performance and a summary of the trigger menu is given. A set of dedicated triggers for the commissioning and calibration of the online $b$-jet selection is also discussed. 

\section{The ATLAS Detector and its Trigger System}
The general-purpose ATLAS detector \cite{AtlasDetector} is built for probing the proton-proton collisions at the LHC. ATLAS satisfies all the requirements needed to successfully operate in the challenging LHC environment and consists of several highly granular and hermetic sub-detectors arranged in concentric layers oriented coaxially with respect to the beam line and centered around the nominal interaction point.

The Inner Detector (ID) provides measurements for high precision tracking and vertexing in the central volume of ATLAS.
The ID consists of three independent subsystems immersed in a 2\,T magnetic field generated by a central solenoid. Closest to the beam line is the pixel detector \cite{PixelDetector} with approximately 80 million silicon pixels of size $50\times400\,\mu$m$^2$ arranged in three layers. A silicon micro-strip detector arranged in four nested cylindrical layers surrounds the pixel detector with a stereoscopic geometry for three-dimensional hit position measurements. The outermost part of the ID consists of layers of straw tubes with 2\,mm radius providing $R-\phi$ measurements out to a radius of about 111\,cm from the beam line. This detector contributes significantly to the track momentum determination and also provides electron identification capabilities through transition radiation measurements.

The ID is designed to reconstruct tracks with a precision of $\mathcal{O}(10\,\mu$m) in the transverse plane, 
to reconstruct masses of unstable particles, to identify the primary vertex among multiple pile-up interactions and to find secondary vertices and displaced tracks to tag jets originating from $b$-quarks ($b$-tagging). 

The ATLAS trigger performs the online event selection in three stages of increasing complexity. The Level-1 trigger (LVL1) is a hardware-based system using coarse information from the calorimeter and dedicated muon trigger chambers. The Level-2 (LVL2) and the final Event Filter (EF) trigger levels are together referred to as the High Level Trigger (HLT). The event selection at both LVL2 and EF is based on software algorithms running on large farms of commercial processors.

LVL2 is the first trigger stage where information from the ID is available. The reconstruction at LVL2 uses specialized fast algorithms to process events in only a few tens of milliseconds on average. At EF, the available processing time is on average a few seconds which enables the use of algorithms developed for the offline reconstruction, but adapted for running in the HLT software framework and tuned for faster execution.

The trigger at LVL1 identifies so called Regions of Interest (RoIs) which are portions of the detector associated to a specific type of physics signature (electron/photon, muon, $\tau$ and jet). RoIs are widely used in the subsequent trigger levels to restrict the amount of data read from the detector readout buffers while accessing the most important part of the event in full granularity.

\section{Online $b$-Tagging Strategy}
The identification of a jet arising from $b$-quark is experimentally possible thanks to several properties: the fragmentation process is hard and the $B$~hadron retains about 70\% of the original $b$-quark momentum, the $B$ hadron mass is relatively high ($>5$\,GeV) and the long $B$-hadron lifetime ($\sim$~1.5\,ps) corresponds to a measurable decay length of $\sim$~3\,mm for a $B$~hadron of $\approx50$\,GeV. Thus, impact parameters of charged tracks and properties of a secondary vertex provide good discriminant power and are widely used as
\begin{wrapfigure}{r}{0.5\textwidth}
\vspace{-20pt}
\begin{center}
\includegraphics[width=0.48\textwidth]{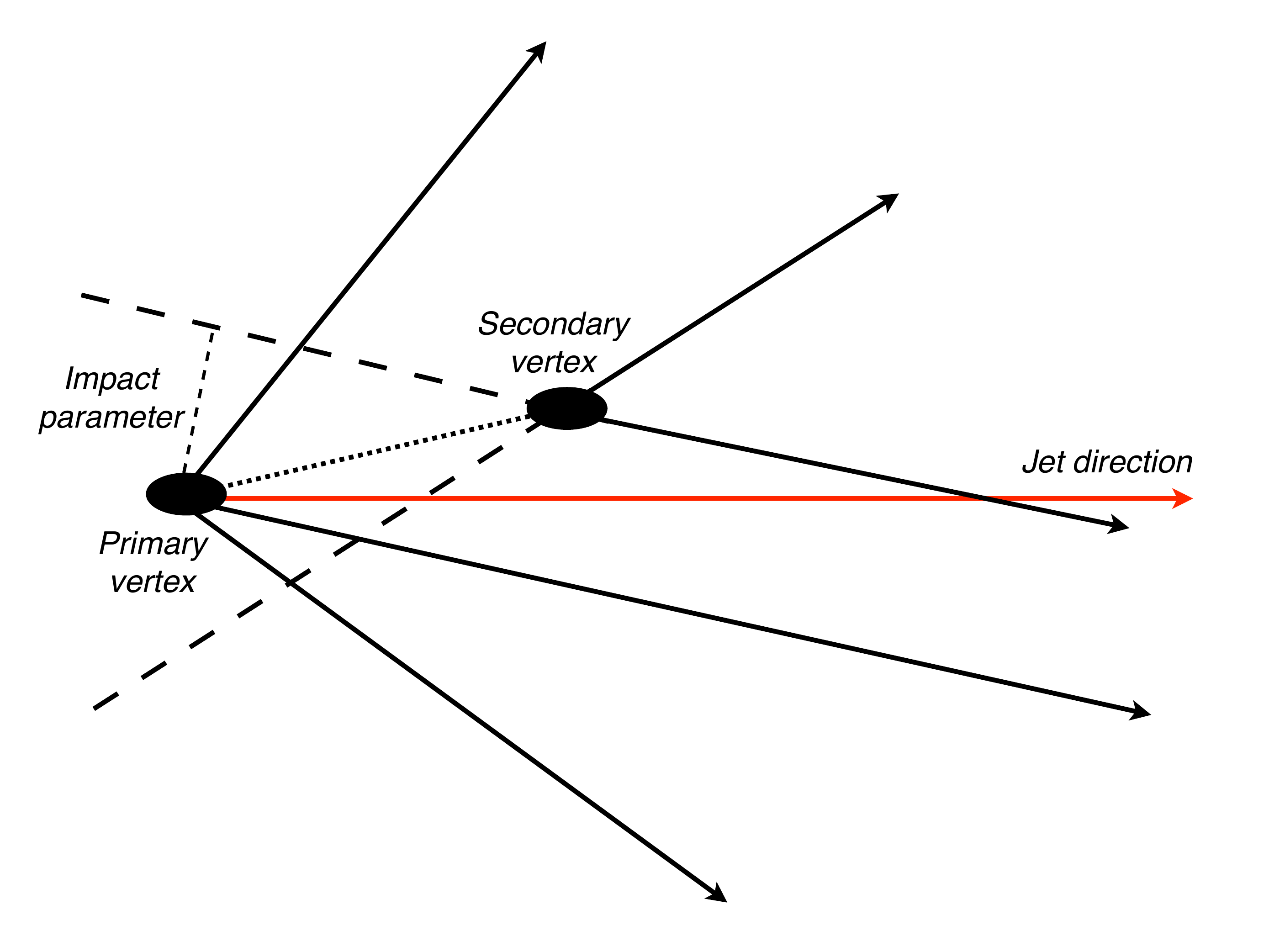}
\end{center}
\vspace{-20pt}
\caption{Schematic sketch (not to scale) of a displaced vertex coming from a $b$-jet with high impact parameter tracks.}
\vspace{-10pt}
\end{wrapfigure} 
input variables for the majority of the $b$-tagging algorithms.

The reconstruction at HLT starts from the RoIs selected by the LVL1 items and the $b$-tagging signatures are activated starting from a jet RoI with dimensions equal to 0.8 in $\eta$ and $\phi$. The LVL1 jet trigger is a fixed-size sliding window algorithm that sums energy in projective towers of size $\Delta\eta\times\Delta\Phi = 0.4\times0.4$. At the HLT, the sequence of algorithms for $b$-jet triggers consists of two major steps: the calorimeter jet reconstruction and the $b$-jet identification. This is a major difference compared to the $b$-jet trigger configuration in 2010 which was exclusively relying on tracking information at the HLT \cite{TriggerPerformance2010} ($b$-jet triggers were not actively selecting events in 2010 due to the relatively low LHC instantaneous luminosity). Each step includes a set of reconstruction algorithms ending with an hypothesis algorithm to stop the trigger execution as soon as it is clear that it doesn't fulfill the trigger requirement anymore.

Three main features must be evaluated to perform the $b$-jet selection: reconstruct tracks of charged particles, measure the primary interaction vertex and estimate the $b$-tagging discriminant variables. This scheme is repeated at both LVL2 and EF.

\subsection{Track Reconstruction}
Online track reconstruction is essential in various trigger signatures (electron, muon, $\tau$ and $b$-jets) and for the online determination of the luminous region in ATLAS. As the various signatures have different requirements, flexible configurations of the pattern recognition are used for ad-hoc optimization purposes. 

At LVL2, two fast custom-built algorithms based on conceptually different techniques, a combinatorial and an histogramming approach, are implemented. In both cases, track candidates are fitted using a fast Kalman filter algorithm. Online operations proved that both approaches at LVL2 can be successfully applied to reconstruct tracks in the harsh tracking environment at the LHC~\cite{TriggerPerformance2010}. Based on slightly different tunings, the primary algorithm adopted for high-multiplicity signatures such as $b$-jets was the combinatorial approach while for high-momentum particles, such as muons, the histogramming approach was the default one. 

At EF, track reconstruction benefits from sharing the majority of the reconstruction software with the offline but slightly adapted to fulfill the trigger requirements of the average event processing time.

\begin{figure}
\begin{center}
\begin{minipage}[t]{0.47\textwidth}
\includegraphics[width=7cm, height=6.5cm]{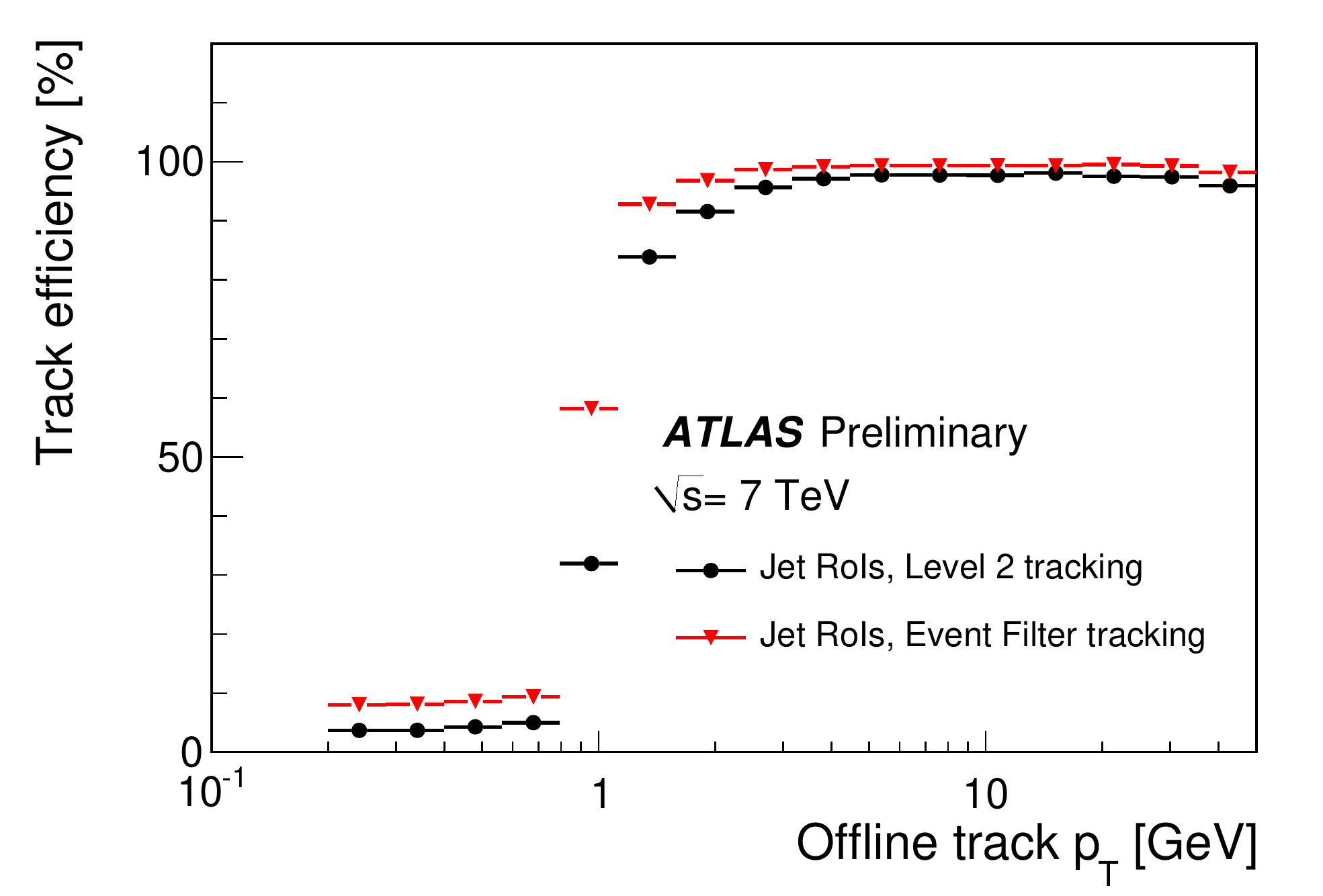}
\caption{Track reconstruction efficiency at LVL2 and EF as a function of the offline track transverse momentum. Online and offline tracks are matched with a geometrical requirement $\Delta R<0.1$ and must fulfill basic quality requirements.}
\label{fig:TrackingPerformance}
\end{minipage}
\,\,\,
\begin{minipage}[t]{0.47\textwidth}
\includegraphics[width=7cm, height=6.3cm]{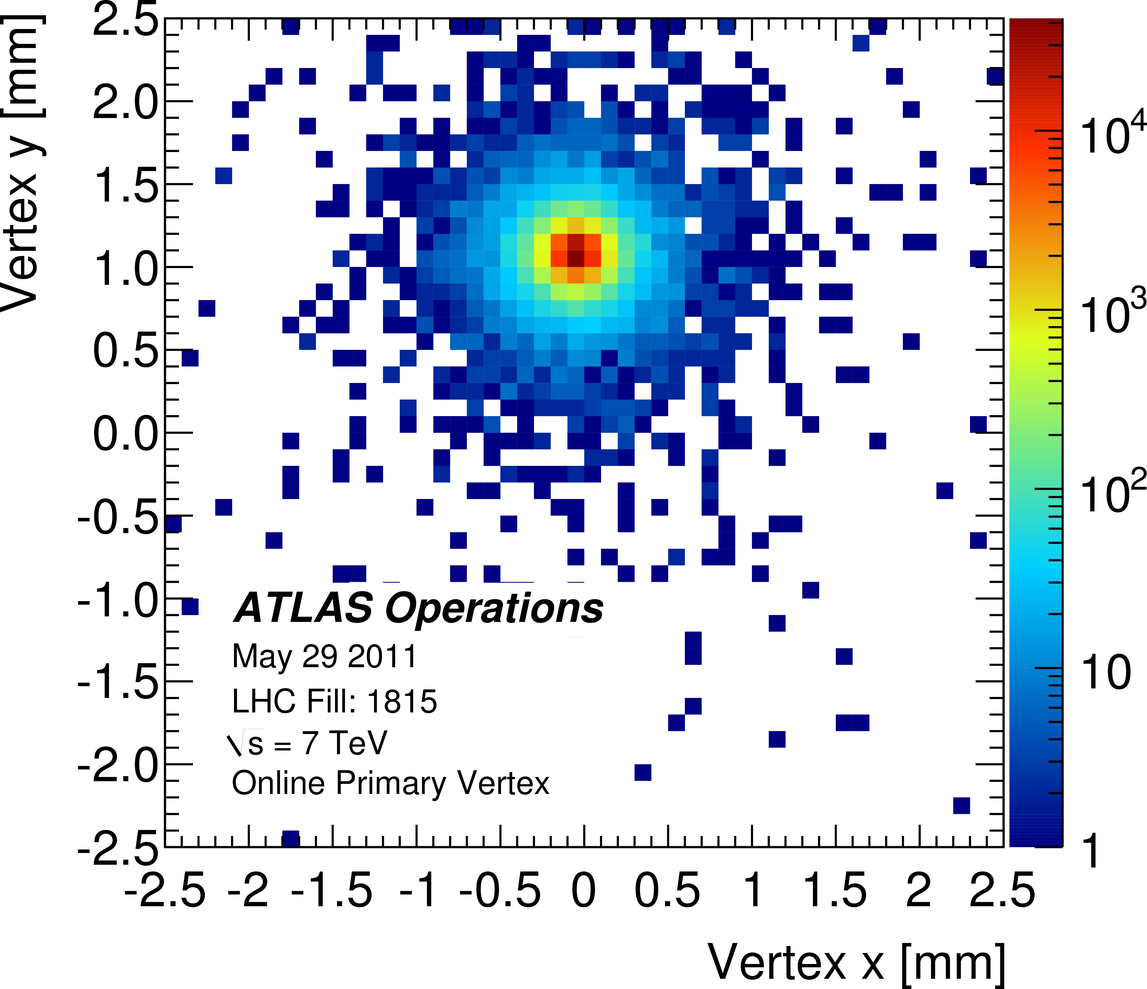}
\caption{The transverse distribution of primary vertices used to derive the luminous region parameters in the trigger framework. The distribution corresponds to 1\,minute of data taking and contains $\mathcal{O}(10^5)$ vertices.}
\label{fig:OnlineBeamSpot}
\end{minipage}
\end{center}
\end{figure}

In order to reduce the processing time, the tracking in $b$-jet triggers is performed in a smaller RoI, half the size in $\eta$ and $\phi$ compared to the corresponding LVL1 jet RoI. A dedicated tuning optimized to efficiently reconstruct tracks with a transverse momentum larger than 1\,GeV and with a transverse impact parameter up to 3\,mm was used in 2011. The track reconstruction efficiency was found to be stable in presence of pile-up and close to 100\% if compared to offline track reconstruction in a wide region of the phase space, as shown in Figure\,\ref{fig:TrackingPerformance}. 

\subsection{Primary Vertex Reconstruction}
The knowledge of the primary vertex position is an essential ingredient to identify $b$-jets. A fast algorithm, exploiting a simple histogramming method based on a sliding window, is found to be highly efficient in finding the true primary vertex position along the beam line with a resolution of about 120 $\mu$m (100 $\mu$m) at LVL2 (EF). 

There is no direct attempt in determining the primary vertex position in the transverse plane as a consequence of reconstructing tracks only in particular RoIs (the primary vertex resolution is strongly correlated with the track multiplicity of the vertex which is relatively low in each RoI, as shown in Figure\,\ref{fig:OnlineBtaggingMultDistribution}). The adopted solution for 2011 was to exploit the online measurement of the luminous region, also called the beamspot, which is estimated in the trigger framework using ad-hoc prescaled triggers. The per-event vertex position is integrated to measure the average position and shape of the beamspot in intervals as short as a few minutes \cite{BeamSpot}. An example of the distribution of primary vertices used in the extraction of the beamspot parameters is shown in Figure\,\ref{fig:OnlineBeamSpot}. 

The commissioning and evaluation of a three-dimensional primary vertex finding algorithm for online $b$-tagging is ongoing and considered for 2012.

\subsection{Discriminating Variable Estimation}
\begin{figure}[t!]
\begin{center}
\subfloat[]{\includegraphics[width=0.33\textwidth]{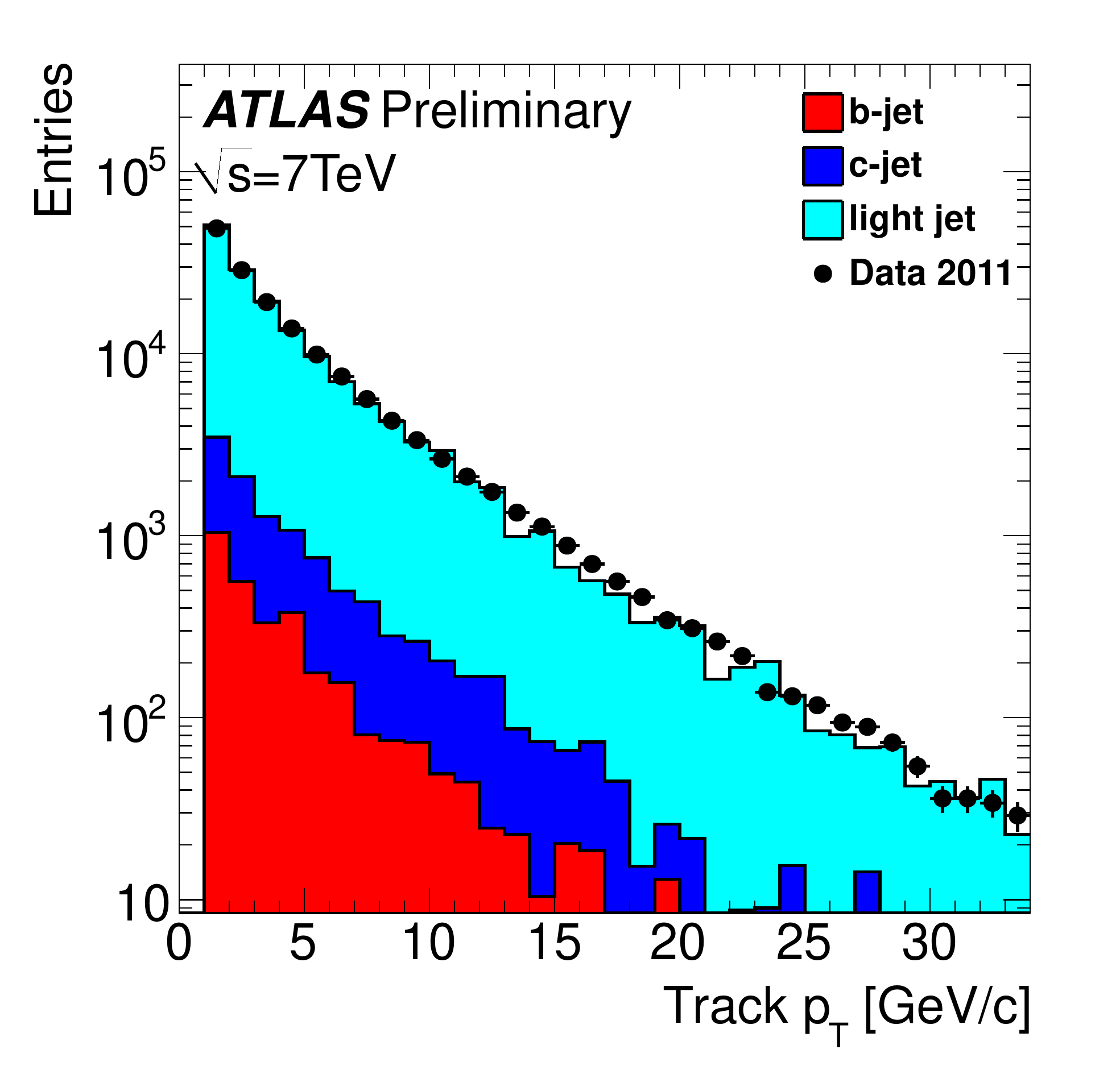}}  
\subfloat[]{\label{fig:OnlineBtaggingMultDistribution}\includegraphics[width=0.33\textwidth]{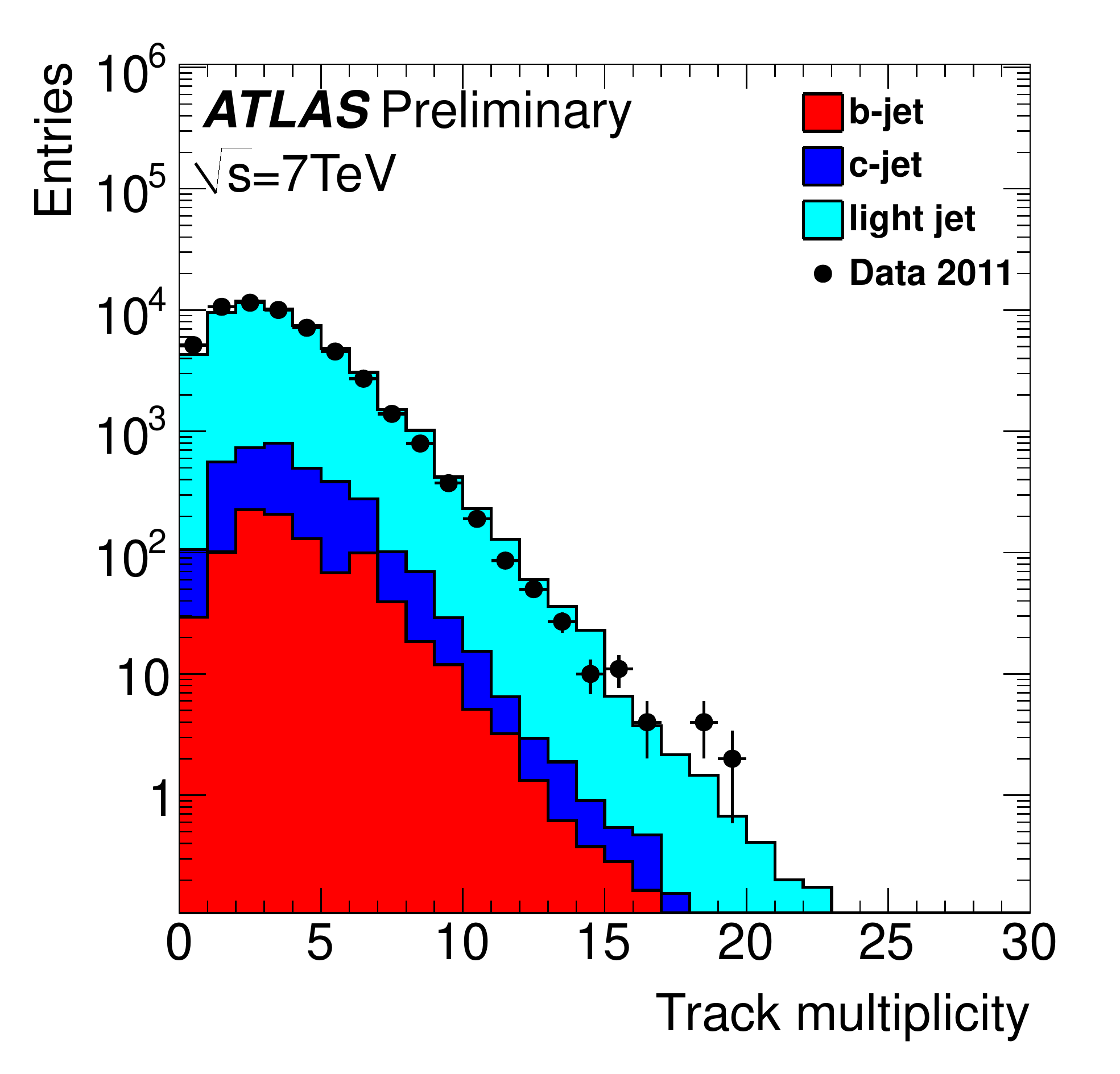}}
\subfloat[]{\label{fig:OnlineBtaggingIPDistribution}\includegraphics[width=0.33\textwidth]{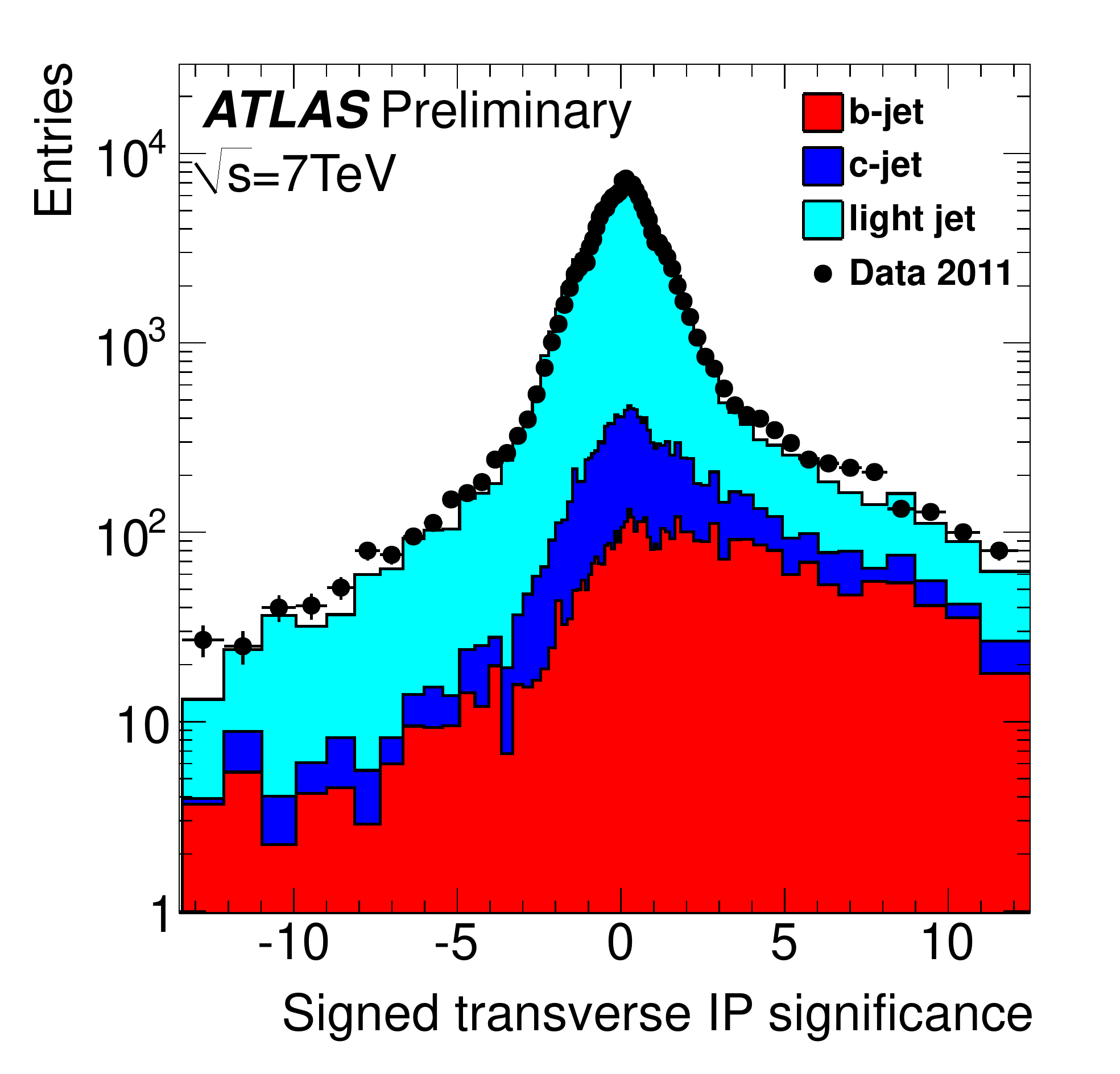}}
\caption{(a) Track transverse momentum (b) track multiplicity and $\mbox{(c)}$ signed transverse impact parameter significance of reconstructed tracks at the EF trigger level. Tracks are reconstructed starting from a low transverse momentum jet identified by the LVL1 and are requested to fulfill online $b$-tagging criteria.}
\label{fig:OnlineBtaggingPerformance}
\,\,\,\,\,\,
\begin{minipage}[t]{0.47\textwidth}
\includegraphics[width=7cm, height=6.3cm]{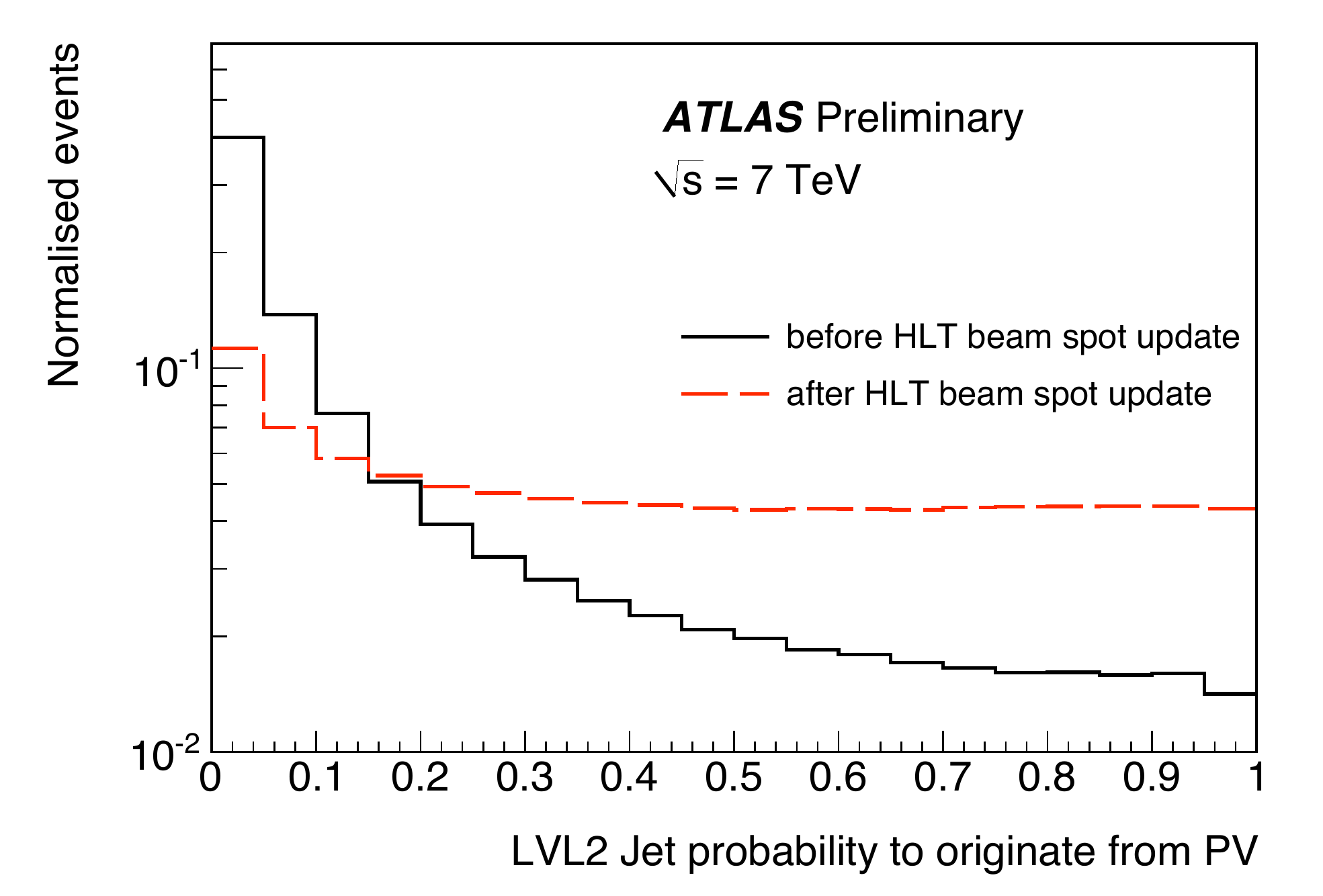}
\caption{Probability for the LVL2 tracks to originate from the primary vertex in case of updated or outdated knowledge of the beam profile in the transverse plane.}
\label{fig:BeamSpotDepedence}
\end{minipage}
\,\,\,
\begin{minipage}[t]{0.47\textwidth}
\includegraphics[width=7cm, height=6.3cm]{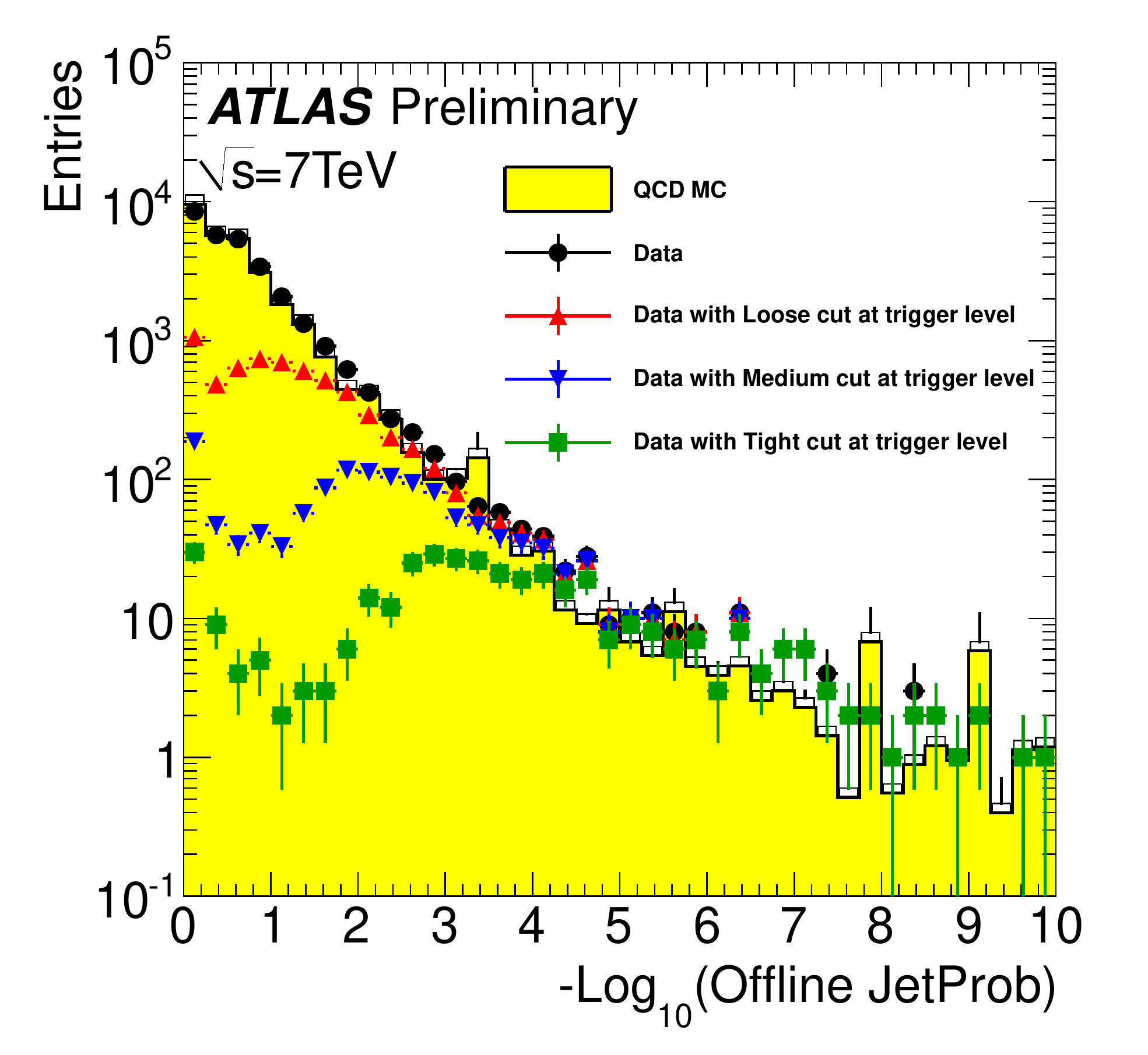}
\caption{The JetProb distribution at the reconstruction level for all jets and jets $b$-tagged at the trigger level with three different operating points.}
\label{fig:OnlineBias}
\end{minipage}
\end{center}
\end{figure}

The online b-tagging in 2011 is based uniquely on the transverse impact parameter of the reconstructed tracks. The signed impact parameter significance is calculated for each track fulfilling specific $b$-tagging selection criteria. The sign is defined using the angle between the jet axis and the line between the primary vertex and the point of closest approach of the track, such that it becomes negative when the track appears to originate behind the primary vertex. As a consequence, most of the tracks produced from decays of particles with a long lifetime, such as $B$ hadrons, are signed as positive, as can be seen in Figure\,\ref{fig:OnlineBtaggingIPDistribution}.

In the significance calculation the transverse impact parameter ($d_0$) is corrected using the beamspot information and its uncertainty using the measured transverse width of the beamspot,
\begin{equation}
\sigma=\sqrt{(\sigma_{d_0}^2+\sigma_{BS}^2)}\,.
\end{equation}

The jet probability method \cite{JetProb}, also called JetProb, is the technique adopted in ATLAS for the online event selection in 2011. 
It computes the probability for a jet to originate from the primary vertex using the signed transverse impact parameter significance of tracks associated to the jet. A probability, $P$, is assigned to each selected track 
\begin{equation}
P=\int_{-\infty}^{-\left|d_0/\sigma(d_0)\right|} \mathcal{R}(x)dx\,,
\end{equation}
where $\mathcal{R}$ is the resolution function for prompt tracks which can be easily measured in data using the negative side of the impact parameter distribution in a light-jet dominated sample. Figure\,\ref{fig:OnlineBtaggingPerformance} shows the signed transverse impact parameters and other track variables at the EF. A good agreement between data and simulation is obtained. In particular, the $b$-jet enrichment in the positive tail of the transverse impact parameter distribution is evident. This enrichment is also enhanced thanks to the employment of a track-jet algorithm which greatly improves the jet direction resolution compared to the jet sliding-window approach at LVL1.

The individual track probabilities, $P_i$, are then combined into a per-jet quantity
\begin{equation}
P_{jet}=P_0\sum_{i=0}^{N_{tracks}-1}\frac{(-\ln P_0)^i}{i!}\,,
\end{equation}
where $P_0=\prod_i P_i$. The distribution of $P_{jet}$ is uniform between zero and one for jets containing tracks compatible with the resolution function $\mathcal{R}$ and is peaked at zero when highly-displaced tracks are reconstructed in the RoI. 

An example of $P_{jet}$ is shown in Figure\,\ref{fig:BeamSpotDepedence} in case of an updated or outdated knowledge of the beamspot parameters. These beamspot parameters are updated for the $b$-jet trigger if the position along any of the three axes changes by more than 10\% or the width along any of the three axes changes by more than 10\% or there is a reduction of 50\% in the statistical uncertainty of any of the values \cite{BeamSpot}. These criteria ensure a flat JetProb distribution for light jets and only jets with JetProb values close to zero, the exact cut being dependent on the algorithm instance, will be contributing in filling the bandwidth allocated to $b$-jet triggers.

Likelihood-ratio methods based on impact parameter significances are also implemented in the trigger domain but did not contribute to the event selection in 2011. The decision to use the JetProb algorithm was driven by its intrinsic robustness, relying only on the knowledge of the negative transverse impact parameter distribution of prompt tracks. Likelihood-ratio methods exploiting both track parameters and secondary vertex properties are being commissioned and considered for 2012.

\begin{figure}[t!]
\begin{center}
\begin{minipage}[t]{0.47\textwidth}
\includegraphics[width=7cm, height=6.3cm]{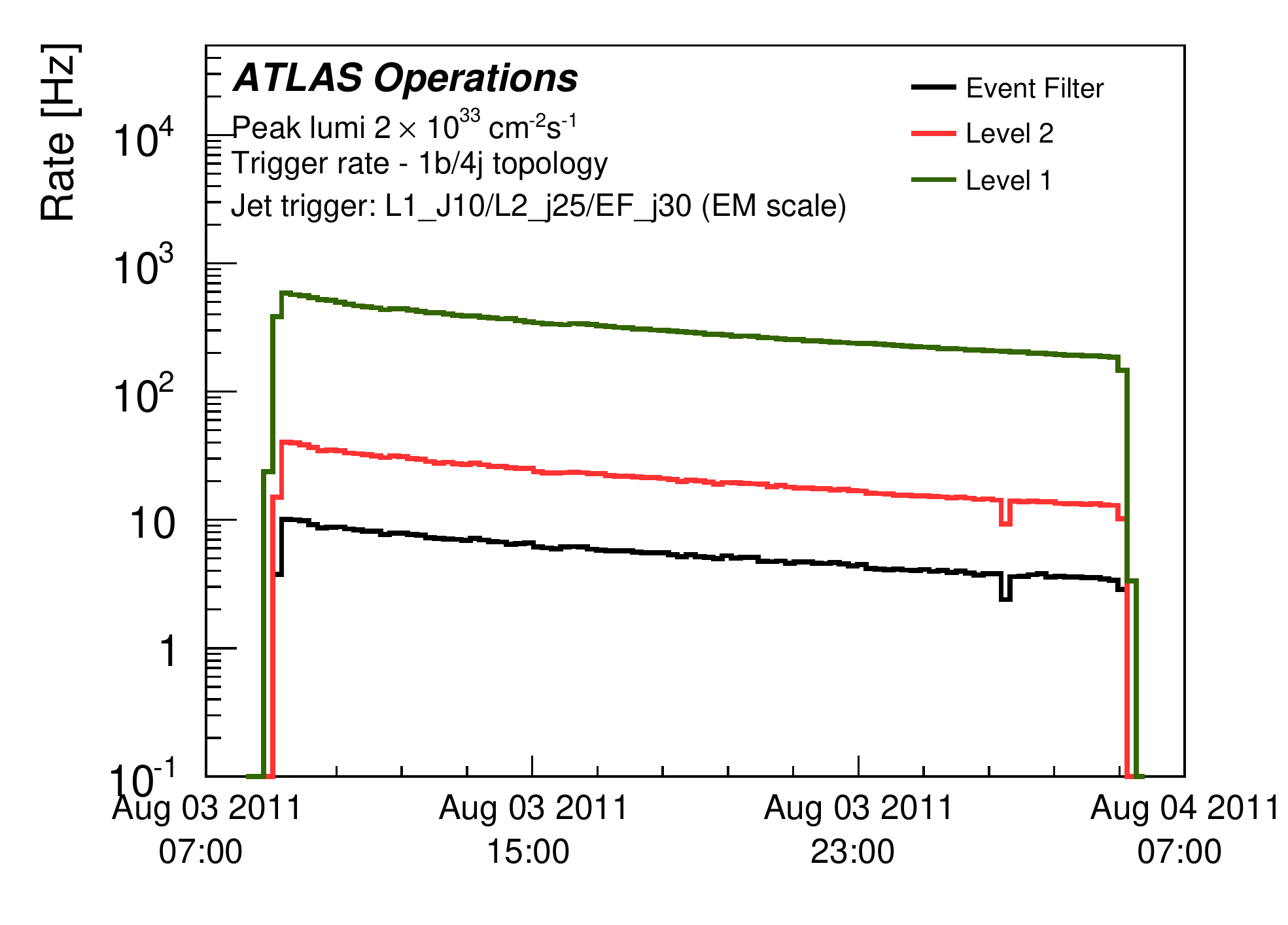}
\caption{Rate of a $b$-jet trigger requiring at least four jets in the event and at least one $b$-tagged jet. The jet thresholds correspond to 10, 25 and 30 GeV at LVL1, LVL2 and EF with energies measured at the electromagnetic scale. The $b$-jet requirement is applied at LVL2 and EF and is tuned to give 60\% efficiency on truth $b$-jets.}
\label{fig:BjetTriggerRate}
\end{minipage}
\,\,\,
\begin{minipage}[t]{0.47\textwidth}
\includegraphics[width=7cm, height=6.3cm]{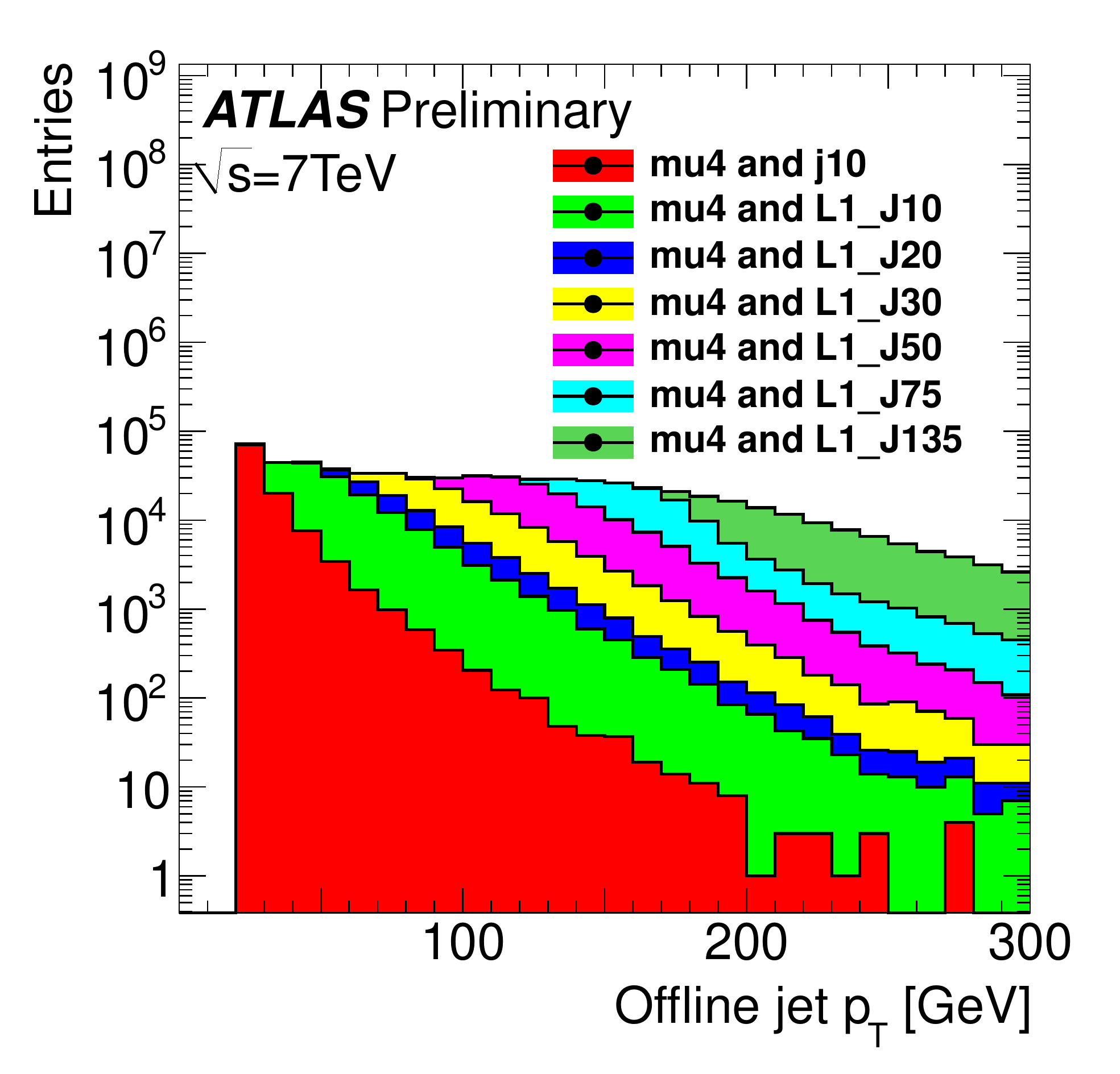}
\caption{Muon-in-jet trigger distribution as a function of the jet transverse momentum in offline muon-in-jet candidates. Several triggers with various jet thresholds select events in order to collect a sample of muon-in-jet candidates in the entire jet transverse momentum spectrum.}
\label{fig:MuJetDistribution}
\end{minipage}
\end{center}
\end{figure}

\section{Trigger Menu and Impact On Offline Quantities}
The primary goal of $b$-jet triggers is to lower jet energy thresholds while keeping the output rate at a manageable level. For 2011, three working points were defined corresponding to 75\%, 60\% and 50\% $b$-jet efficiency on a simulated top quark sample, called the loose, medium and tight instances of the $b$-tagging algorithm. The medium instance was the most widely used in $b$-jet triggers for physics analyses.

Figure\,\ref{fig:OnlineBias} shows the bias on the offline JetProb weight distribution coming from the online $b$-tagging. Data and simulation are overlaid in case of no $b$-jet trigger requirement and the effect on data when a $b$-jet trigger requirement is included produces a bias on the offline distribution. Offline $b$-jet candidates are however recorded by the trigger system, a clear indication of the correlation between the online and offline taggers.

The $b$-jet trigger rejection is evident from Figure\,\ref{fig:BjetTriggerRate} where the rate at the output of the three trigger levels is shown for a particular $b$-jet trigger in a run with peak instantaneous luminosity of $2\times10^{33}$\,cm$^{-2}$s$^{-1}$. This particular trigger was designed for the all-hadronic $t\overline{t}$ final state. Other $b$-jet triggers were implemented and running online for benchmark channels such as $H\rightarrow b\overline{b}$, especially useful for the low Higgs mass region still to be fully explored, and for supersymmetric models with no leptons in the final state.

The combination of the online plus offline $b$-tagging efficiency and mis-tag rate is a key ingredient to fully integrate $b$-jet triggers into physics analyses. Since detailed tracking information is not perfectly reproduced by the simulation, these measurements are currently being derived with data-driven techniques and are carried out using analyses already employed in ATLAS for the offline $b$-tagging \cite{BtaggingCalibration}. These methods mainly explore muon properties from heavy-flavour decays or explicit reconstruction of $B$ hadron decay chains. To successfully apply these methods a set of dedicated triggers collected muons close to jets in 2011, obtaining a sample enriched in $B$~hadrons with a muon in the decay chain. Triggers with different jet thresholds were used to obtain a rather flat jet transverse momentum spectrum, as clearly shown in Figure\,\ref{fig:MuJetDistribution}. In the second part of 2011, an additional trigger was deployed online to enable the $b$-jet efficiency measurement using techniques based on the reconstruction of the top quark.

\section{Conclusions and Outlook}
Exploiting $b$-jet identification at the trigger level allows for lowering jet trigger thresholds. $b$-jet triggers are a reality in ATLAS since the beginning of 2011. Various triggers were used to collect events for a set of benchmarks channels and contributed to maintain an acceptable output rate. Online tracking is an essential ingredient for the $b$-jet selection and has shown high performance and robustness over a large range of working conditions.

Physics analyses relying on $b$-jet triggers are in progress and will benefit from the ongoing measurements of the online and offline $b$-tagging efficiency and mis-tag rate using data-driven techniques.

\end{document}